\documentclass[11pt]{article}
\def\half{\frac{1}{2}}

\def\pl{ \: + }
\def\mi{ \: - }

\newfont{\bbbold}{msbm10 scaled \magstep1}

\def\bbR{\mbox{\bbbold R}}

\def\cR{{\cal R}}

\newfont{\goth}{eufm10 scaled \magstep1}

\def\ge{\mbox{\goth e}}

\def\gg{\mbox{\goth g}}

\def\go{\mbox{\goth o}}

\def\gR{\mbox{\goth R}}
\def\gs{\mbox{\goth s}}

\def\gu{\mbox{\goth u}}

\def\a{\alpha}

\def\C{\Gamma}
\def\d{\delta}\def\D{\Delta}

\def\f{\phi}\def\F{\Phi}\def\vf{\varphi}
\def\h{\eta}

\def\l{\lambda}

\def\p{\pi}
\def\P{\Pi}

\def\S{\Sigma}

\def\th{\theta}

\def\be{\begin{equation}}\def\ee{\end{equation}}
\def\bea{\begin{eqnarray}}\def\eea{\end{eqnarray}}
\def\barr{\begin{array}}\def\earr{\end{array}}

\def\o{\omega}\def\O{\Omega}
\def\del{\partial}

\def\xz{\times}

\def\nab{\nabla}

\def\hE{\hat{E}}\def\hR{\hat{R}}

\def\tnab{\tilde{\nabla}}

\def\hE{\widehat{E}}

\def\hL{\widehat{L}}
\def\hR{\widehat{R}}

\def\ghh{\widehat{\h}}
\def\ghl{\widehat{\l}}

\def\hnab{\widehat{\nab}}

\def\pl{{(+)}}\def\mi{{(-)}} \def\plmi{{(\pm)}}
\let\la=\label

\def\nn{\nonumber}
\def\bd{\begin{document}}
\def\ed{\end{document}}
\def\ba{\begin{array}}
\def\ea{\end{array}}
\def\bea{\begin{eqnarray}}
\def\eea{\end{eqnarray}}
\def\ft#1#2{{\textstyle{{\scriptstyle #1}\over {\scriptstyle #2}}}}
\def\fft#1#2{{#1 \over #2}}
\newcommand{\eq}[1]{(\ref{#1})}
\newcommand{\w}[1]{\\[0.#1cm]}
\def\eqs#1#2{(\ref{#1}-\ref{#2})}
\def\det{{\rm det\,}}
\def\tr{{\rm tr}}
\newcommand{\hoch}[1]{$\, ^{#1}$}
\newcommand{\tamphys}{\it\small Center for Theoretical Physics,
Texas A\&M University, College Station, TX 77843, USA}
\newcommand{\kings}
{\it\small Department of Mathematics, King's College, London, UK}
\newcommand{\uu}
{\it\small Department of Theoretical Physics, Uppsala, Sweden}
\newcommand{\hip}
{\it\small HIP-Helsinki Institute of Physics, P.O. Box 64 FIN-00014
University of Helsinki, Suomi-Finland}
\newcommand{\stock}
{\it\small Department of Theoretical Physics, Stockholm, Sweden}
\makeatletter
\renewcommand\theequation{\thesection.\arabic{equation}}
\@addtoreset{equation}{section} \makeatother

\newcommand{\auth}
{\large P.S. Howe\hoch{1,2}, U. Lindstr\"om\hoch{2,3} and V.
Stojevic \hoch{1}}

\begin{document}

\hfill{KCL-TH-05-08}

\hfill{UUITP-10/05}

\hfill{HIP-2005-28/TH}

\hfill{hep-th/0507035}

\hfill{\today}

\vspace{20pt}

\begin{center}
{\Large{\bf Special holonomy sigma models with boundaries}}
\vspace{30pt}

\auth

\vspace{15pt}

\begin{itemize}
\item [$^1$] \kings \item [$^2$] \uu \item  [$^3$] \hip
\end{itemize}

\vspace{60pt}

{\bf Abstract}

\end{center}

A study of $(1,1)$ supersymmetric two-dimensional non-linear sigma
models with boundary on special holonomy target spaces is presented.
In particular, the consistency of the boundary conditions under the
various symmetries is studied. Models both with and without torsion
are discussed.

\pagebreak \tableofcontents \setcounter{page}{1}


\section{Introduction}


There has been a long history of interplay between differential
geometry and supersymmetric non-linear sigma models starting with
the observation that $N=2$ supersymmtery in two dimensions requires
the sigma model target space to be a K\"ahler manifold
\cite{Zumino:1979et}. It was first pointed out in
\cite{Delius:1989fy} that one could construct conserved currents in
$(1,1)$ sigma models given a covariantly constant form on the target
space, and in \cite{Odake:1988bh} it was shown that the $(1,1)$
model on a Calabi-Yau three-fold has an extended superconformal
algebra involving precisely such a current constructed from the
holomorphic three-form. In \cite{Howe:1991ic} symmetries of this
type were studied systematically in the classical sigma model
setting; each manifold on Berger's list of irreducible non-symmetric
Riemannian manifolds has one or more covariantly constant forms
which give rise to conserved currents and  the corresponding Poisson
bracket algebras are non-linear, i.e. they are of W-symmetry type.
Subsequently the properties of these algebras were studied more
abstractly in a conformal field theory framework
\cite{Shatashvili:1994zw,Figueroa-O'Farrill:1996hm} and more
recently in topological models \cite{deBoer:2005pt}.

In this paper we shall discuss two-dimensional (1,1) supersymmetric
sigma models with boundaries with extra symmetries of the above
type, focusing in particular on target spaces with special holonomy.
In a series of papers
\cite{Albertsson:2001dv}-\cite{Albertsson:2003va} classical
supersymmetric sigma models with boundaries have been discussed in
detail and it has been shown how the fermionic boundary conditions
involve a locally defined tensor $R$ which determines the geometry
associated with the boundary. In particular, in the absence of
torsion, one finds that there are integral submanifolds of the
projector $P=\half (1+R)$ which have the interpretation of being
branes where the boundary can be located. These papers considered
$(1,1)$ and $(2,2)$ models and the analysis was also extended to
models of this type with torsion where the intepretation of $R$ is
less straightforward. The main purpose of the current paper is to
further extend this analysis to include symmetries associated with
certain holonomy groups or $G$-structures. We shall discuss models
both with and without torsion.

Torsion-free sigma models with boundaries on manifolds with special
holonomy were first considered in \cite{Becker:1996ay} where it was
shown how the identification of the left and right currents on the
boundary has a natural interpretation in terms of calibrations and
calibrated submanifolds. Branes have also been discussed extensively
in boundary CFT \cite{Schomerus:2002dc}, including the $G_2$ case
\cite{Roiban:2001cp}, and in topological string theory
\cite{Bershadsky:1995qy}.

The main new results of the paper concern boundary  (1,1) models
with torsion or with a gauge field on the brane. There is no
analogue of Berger's list in the case of torsion but we can
nevertheless consider target spaces with specific $G$-structures
which arise due to the presence of covariantly constant forms of the
same type. In order to generalise the discussion from the
torsion-free case we require there to be two independent
$G$-structures specified by two sets of covariantly constant forms
$\{\l^+,\l^-\}$ which are covariantly constant with respect to two
metric connections $\{\C^+,\C^-\}$ and which have closed
skew-symmetric torsion tensors $T^{\pm}=\pm H$, where $H=d b$, $b$
being the two-form potential which appears in the sigma model
action.  This sort of structure naturally generalises the notion of
bi-hermitian geometry which occurs in $N=2$ sigma models with
torsion \cite{Gates:1984nk,Howe:1991ic} and which has been studied
in the boundary sigma model context in \cite{Lindstrom:2002jb}. We
shall refer to this type of structure as a bi-$G$-structure. The
groups $G$ which are of most interest from the point of view of
spacetime symmetry are the groups which appear on Berger's list and
for this reason we use the term special holonomy. Bi-$G$-structures
are closely related to the generalised structures which have
appeared in the mathematical literature
\cite{Hitchin:2004ut,Gualtieri,Witt:2004vr}. These generalised
geometries have been discussed in the $N=2$ sigma model context
\cite{Lindstrom:2004iw,Kapustin:2004gv,Zabzine:2004dp}. In a recent
paper they have been exploited in the context of branes and
generalised calibrations.

We shall show that, in general, the geometrical conditions implied
by equating the left and right currents on the boundary lead to
further constraints by differentiation and that these constraints
are the same as those which arise when one looks at the stability of
the boundary conditions under symmetry transformations. It turns
out, however, that these constraints are automatically satisfied by
virtue of the target space geometry.

We then study the target space geometry of some examples, in
particular bi-$G_2$, bi-$SU(3)$ and bi-$Spin(7)$ structures.
Structures of this type have appeared in the supergravity literature
in the context of supersymmetric solutions with flux
\cite{Gauntlett:2002sc,Grana:2004bg,Grana:2005ny}.

The paper is organised as follows: in section two we review the
basics of boundary sigma models, in section three we discuss
additional symmetries associated with special holonomy groups or
bi-$G$-structures, in section four we examine the consistency of the
boundary conditions under symmetry variations, in section five we
look at the target space geometry of bi-$G$ structures from a simple
point of view and in section six we look at some examples of
solutions of the boundary conditions for the currents defined by the
covariantly constant forms.


\section{Review of basics}


The action for a $(1,1)$-supersymmetric  sigma model without
boundary is

\begin{equation}
S=\int\, dz\, e_{ij} D_+ X^i D_- X^j\ , \label{2.1}
\end{equation}

where

\begin{equation}
e_{ij}:=g_{ij}+b_{ij}\ , \label{2.2}
\end{equation}

$b$ being a two-form potential with field strength $H=db$ on the
$n$-dimensional Riemannian target space $(M,g)$. $X^i, i=1,\ldots
n$, is the sigma model field represented in some local chart for $M$
and $z$ denotes the coordinates of $(1,1)$ superspace $\S$. We shall
use a light-cone basis so that $z=(x^{++},x^{--},\th^{+},\th^{-})$,
with $x^{++}=x^0+x^1, x^{--}=x^0- x^1$. $D_+$ and $D_-$ are the
usual flat superspace covariant derivatives which obey the relations

\begin{equation}
D_+^2 = i\del_{++};\qquad D_-^2 = i\del_{--};\qquad \{D_+,D_-\}= 0 \
. \label{2.3}
\end{equation}

We use the convention that $\del_{++} x^{++}=1$. We shall take the
superspace measure to be

\begin{equation}
dz:= d^2 x\,D_+ D_- \label{2.4}
\end{equation}

with the understanding that the superfield obtained after
integrating over the odd variables (i.e after applying $D_+ D_-$ to
the integrand) is to be evaluated at $\th=0$.

As well as the usual Levi-Civita connection $\nab$ there are two
natural metric connections $\nab^{\pm}$ with torsion
\cite{Gates:1984nk,Howe:1985pm},

\begin{equation}
\C^{(\pm)}{}^j_{ik}:=\C_{ik}^{j}\pm \half H^j{}_{ik}\ . \label{2.5}
\end{equation}

The torsion tensors of the two connections are given by

\begin{equation}
T^{(\pm)}{}^i_{jk}=\pm H^i{}_{jk}\ , \label{2.6}
\end{equation}

so that the torsion is a closed three-form in either case.

In the presence of a boundary, $\del\S$, it is necessary to add
additional boundary terms to the action \eq{2.2} when there is
torsion \cite{Albertsson:2002qc}. The boundary action is

\begin{equation}
S_{bdry}= \int_{\del\S}\, a_i \dot X^i +\frac{i}{4}b_{ij}(\psi_+^i
\psi_+^j +\psi_-^i\psi_-^j) \label{2.6.1}\ ,
\end{equation}

where $a_i$ is a gauge field  which is defined only on the
submanifold where the boundary sigma model field maps takes its
values. Note that the boundary here is purely bosonic so that the
fields are component fields, $\psi_{\pm}^i:=D_{\pm} X^i|$, the
vertical bar denoting the evaluation of a superfield at
$\th=0$).\footnote{We shall use $X^i$ to mean either the superfield
or its leading component; it should be clear from the context which
is meant.} The boundary term ensures that the action is unchanged if
we add $d c$ to $b$ provided that we shift $a$ to $a-c$. The
modified field strength $F=f +b$, where $f=da$,  is invariant under
this transformation. In the absence of a $b$-field one can still
have a gauge field on the boundary.

In the following we briefly summarise the approach to boundary sigma
models of references
\cite{Albertsson:2001dv}-\cite{Albertsson:2003va}. We impose the
standard boundary conditions \cite{Callan:1988wz} on the fermions,

\begin{equation}
\psi_-^i =\h R^i{}_j \psi_+^j, \qquad \h=\pm1, \qquad  {\rm on}\
\del\S \label{2.7}
\end{equation}

We shall also suppose that there are both Dirichlet and Neumann
directions for the bosons. That is, we assume that there is a
projection operator $Q$ such that

\begin{equation}
Q^i{}_j \d X^j = Q^i{}_j \dot X^j=0 \ , \label{2.7.2}
\end{equation}

on $\del\S$. If $F=0$, parity implies that $R^2=1$, so that $Q=\half
(1-R)$, while $P:=\half (1+R)$ is the complementary projector. In
general, we shall still use $P$ to denote $\half (1+R)$ and the
complementary projector will be denoted by $\p,\ \p:=1-Q$. We can
take $Q$ and $\p$ to be orthogonal

\begin{equation}
\p^k_i g_{kl} Q^l{}_j=0 \ . \label{2.8}
\end{equation}

Equation \eq{2.7.2} must hold for any variation along the boundary.
Making a supersymmetry transformation we find

\begin{equation}
QR + Q=0\ . \label{2.9}
\end{equation}

On the other hand, the cancellation of the fermionic terms in the
boundary variation (of $S+ S_{bdry}$), when the bulk equations of
motion are satisfied, requires

\begin{equation}
g_{ij}=g_{kl} R^k{}_i R^l{}_j \ . \label{2.18}
\end{equation}

Using this together with orthogonality one deduces the following
algebraic relations,

\begin{eqnarray}
QR &=& RQ= -Q;\qquad QP= PQ =0;\nn \w2 \p P&=& P \p =P;\qquad \ \ \
\p R= R \p \ . \label{2.10}
\end{eqnarray}

Making a supersymmetry variation of the fermionic boundary condition
\eq{2.7} and using the equation of motion for the auxiliary field,
$F^i:=\nab^\pl_- D_+ X^i|$, namely $F^i=0$, we find the bosonic
boundary condition\footnote{The occurrence of (combinations of)
field equations as boundary conditions is discussed in
\cite{Lindstrom:2002mc}.}

\begin{equation}
i(\del_{--} X^i - R^i{}_j \del_{++} X^j) = (2 \tnab_j R^i{}_k
-P^i{}_l H^l_{jm}R^m{}_k) \psi_{+}^j\psi_+^k \ , \label{2.11}
\end{equation}

where $\tnab$ is defined by

\begin{equation}
\tnab_i:=P^j{}_i \nab_j\ . \label{2.12}
\end{equation}

Combining \eq{2.11} with the bosonic boundary condition arising
directly from the variation we find

\begin{equation}
\hE_{ji} = \hE_{ik} R^k {}_j\ , \label{2.12.1}
\end{equation}

where

\begin{equation}
E_{ij}:= g_{ij} + F_{ij} \label{2.12.2}
\end{equation}

and the hats denote a pull-back to the brane,

\begin{equation}
\hE_{ij}:=\p^k{}_i \p^l{}_j E_{kl}\ . \label{2.12.3}
\end{equation}

From \eq{2.12.3} we find an expression for $R$,

\begin{equation}
R^i{}_j=(\hE^{-1})^{ik} \hE_{jk}- Q^i{}_j \ , \label{2.12.4}
\end{equation}

where the inverse is taken in the tangent space to the brane, i.e.

\begin{equation}
(\hE^{-1})^{ik} \hE_{kj}=\p^i{}_j\ . \label{2.12.5}
\end{equation}

We can multiply equation \eq{2.11} with $Q$ to obtain

\begin{equation}
P^l{}_{[i} P^m{}_{j]}\nab_l Q^k{}_m=0\ . \label{2.13}
\end{equation}

Using \eq{2.10} we can show that this implies  the integrability
condition for $\p$,

\begin{equation}
\p^l{}_{[i} \p^m{}_{j]}\nab_l Q^k{}_m=0\ . \label{2.14}
\end{equation}

This confirms that the distribution specified by $\p$ in $TM$ is
integrable and the boundary maps to a submanifold, or brane, $B$.
However, in the Lagrangian approach adopted here, this is implicit
in the assumption of Dirichlet boundary conditions. When $F=0$ the
derivative of $R$ along the brane is essentially the second
fundamental form, $K$. Explicitly,

\begin{equation}
K^i_{jk}=P^l{}_{j} P^m{}_{k}\nab_l Q^i{}_m = P^l{}_j\tnab_k Q^i{}_l
. \label{2.14.1}
\end{equation}

The left and right supercurrents are

\begin{eqnarray}
T_{+3}:&=&g_{ij} \del_{++}X^i D_+ X^j -\frac{i}{6} H_{ijk} D_{+3}
X^{ijk}\w1 T_{-3}:&=&g_{ij} \del_{--}X^i D_- X^j +\frac{i}{6}
H_{ijk} D_{-3} X^{ijk} \label{2.15}
\end{eqnarray}

The conservation conditions are

\begin{equation}
D_-T_{+3}=D_+ T_{-3}=0 \ . \label{2.16}
\end{equation}

The superpartners of the supercurrents are the left and right
components of the energy-momentum tensor, $D_+ T_{+3}$ and $D_-
T_{-3}$ respectively. If one demands invariance of the total action
under supersymmetry one finds that, on the boundary, the currents
are related by

\begin{eqnarray}
T_{+3} &=& \h T_{-3}\label{2.16.1}\w1 D_+ T_{+3}&=& D_- T_{-3}\ .
\label{2.17}
\end{eqnarray}

The supercurrent boundary condition has a three-fermion term which
implies the vanishing of the totally antisymmetric part of

\begin{equation}
2Y_{i,jk} + P^l{}_i H_{ljm} R^m{}_k +\frac{1}{6}(H_{ijk} + H_{lmn}
R^l{}_i R^m{}_j R^n{}_k)\ , \label{2.19}
\end{equation}

where

\begin{equation}
Y_{i,jk}:= (R^{-1})_{jl} \tnab_i R^l{}j\ . \label{2.20}
\end{equation}


\section{Additional symmetries}


A general variation of \eq{2.1}, neglecting boundary terms, gives

\begin{eqnarray}
\d S&=&\int\,dz\, 2g_{ij}\d X^i \nab^{(+)}_- D_+ X^j \nn \w1
&=&-\int\,dz\, 2g_{ij}\d X^i g_{ij} \nab^{(-)}_+ D_- X^j \ .
\label{3.1}
\end{eqnarray}

The additional symmetries we shall discuss are transformations of
the form,

\begin{equation}
\d_{\pm} X^i= a^{\pm \ell} L^{(\pm)}{}^i{}_{j_1\ldots j_{\ell}}
D_{\pm \ell} X^{j_1\ldots j_{\ell}}\ ,\qquad  D_{\pm \ell}
X^{j_1\ldots j_{\ell}}:= D_{\pm} X^{j_1}\ldots D{\pm}+ X^{j_{\ell}}\
, \label{3.2}
\end{equation}

where $L^{(\pm)}$ are vector-valued ${\ell}$-forms such that

\begin{equation}
\l^{\plmi}{}_{i_1\ldots i_{{\ell}+1}}:= g_{i_1 j} L^{\plmi
j}{}_{i_2\ldots i_{{\ell}+1}} \label{3.3}
\end{equation}

are $(\ell+1)$-forms which are covariantly constant with respect to
$\nab^{(\pm)}$. For example,  a left transformation of this type
gives

\begin{eqnarray}
\d S&=&\int\,dz\, 2 a^{+\ell}\l^{\pl}{}_{i_1\ldots i_{{\ell}+1}}
D_{+\ell}X^{i_2\ldots i_{\ell+1}} \nab^{(+)}_- D_+ X^{i_1} \nn \w1
&=&\int\,dz\,\frac{2}{\ell+1}a^{+\ell}\l^{\pl}{}_{i_1\ldots
i_{{\ell}+1}} \nab^{(+)}_- D_{+(\ell+1)} X^{i_1\ldots i_{\ell+1}}
\nn \w1 &=&\int\,dz\, (-1)^{\ell}
D_-\left(\frac{2}{\ell+1}a^{+\ell}\l^{\pl}{}_{i_1\ldots
i_{{\ell}+1}} D_{+(\ell+1)} X^{i_1\ldots i_{\ell+1}}\right) \ ,
\label{3.4}
\end{eqnarray}

where the last step follows from covariant constancy of $\l^\pl$ and
the chirality of the parameters,

\begin{equation}
D_- a^{+\ell}=D_+ a^{-\ell}=0\ . \label{3.5}
\end{equation}

Hence these transformations are symmetries of the sigma model
without boundary. In the torsion-free case the $\l$s will be the
forms which exist on the non-symmetric Riemannian manifolds on
Berger's list. There is no such list in the presence of torsion but
the same forms will define reductions of the structure group to the
various special holonomy groups. In order to preserve the symmetry
on the boundary we must have both left and right symmetries so there
must be two independent such reductions. Thus we can say that we are
interested in boundary sigma models on manifolds which have
bi-$G$-structures.

The $\l$-forms can be used to construct currents $L^{(\pm)}_{\pm
(\ell +1)}$,

\begin{equation}
L^{(\pm)}_{\pm (\ell +1)}:=\l^{(\pm)}{}_{i_1\ldots i_{{\ell}+1}}
D_{\pm(\ell+1)} X^{i_1\ldots i_{\ell+1}} \label{3.6}
\end{equation}

If we make both left and right transformations of the type \eq{3.2}
we obtain

\begin{eqnarray}
\d S&=&\frac{2(-1)^{\ell}}{\ell+1}\int\, d^2 x D_+ D_-\,
\left(D_-(a^{+\ell}L^{(+)}_{\pm (\ell +1)})- D_+(a^{-\ell}L^{-}_{-
(\ell +1)})\right) \nn\w2
&=&\frac{i(-1)^{\ell+1}}{\ell+1}\int_{\del\S}\,
\left(D_+(a^{+\ell}L^{(+)}_{\pm (\ell +1)})- D_-(a^{-\ell}L^{-}_{-
(\ell +1)})\right) \ . \label{3.6.1}
\end{eqnarray}

In order for a linear combination of the left and right symmetries
to be preserved in the presence of a boundary the parameters should
be related by

\begin{eqnarray}
a^{+{\ell}} &=& \h_L a^{-{\ell}}\ , \w1 D_+ a^{+{\ell}} &=& \h \h_L
D_-a^{-{\ell}}\ , \label{3.9}
\end{eqnarray}

on the boundary, where $\h_L=\pm 1$.\footnote{In the case that there
is one pair of $L$ tensors.} This implies that the currents and
their superpartners should satisfy the boundary conditions

\begin{eqnarray}
L^{(+)}_{+({\ell}+1)}&=& \h\h_L L^{(-)}_{-({\ell}+1)}\label{3.7}\
,\w1 D_+ L^{(+)}_{+({\ell}+1)}&=& \h_L D_- L^{(-)}_{-({\ell}+1)}\ .
\label{3.8}
\end{eqnarray}

The boundary condition \eq{3.7} implies

\begin{equation}
\l^{(+)}{}_{i_1\ldots i_{{\ell}+1}}=\h_L \h^{\ell}
\l^{(-)}{}_{j_1\ldots j_{{\ell}+1}}R^{j_1}{}_{i_1}\ldots
R^{j_{{\ell}+1}}{}_{i_{{\ell}+1}}\ . \label{3.10}
\end{equation}

The algebra of left (or right) transformations was computed in the
torsion-free case in \cite{Howe:1991ic}. The commutators involve
various generalised Nijenhuis tensors and the classical algebra has
a non-linear structure of $W$-type. In fact, the generalised
Nijenhuis tensors vanish in the absence of torsion. However, this is
not the case when torsion is present. The commutator of two plus
transformations of the type given in \eq{3.2} is (we drop the pluses
on the tensors to simplify matters),

 \begin{equation}
 [\d_L,d_M]=\d_P + \d_N + \d_K
 \la{3.10.1}
 \end{equation}

where $P$ and $N$ are antisymmetric tensors given by

 \be
 P_{LM}= (L\cdot M)_{[L,M]}:=L_{p[L} M^p{}_{M]}
 \la{3.10.2}
 \ee

and

 \be
 N_{iLM}=(\ell+m+1) H_{jk[i} L^j{}_{L} M^k{}_{M]}+(-1)^{\ell}
 \frac{\ell m}{6}H_{[i\ell_1\ell_2}Q_{L_3 M]}\ .
 \la{3.10.3}
 \ee

The $(\ell+m-2)$-form $Q$ is defined by

 \be
 Q_{L_2 M_2}= \frac{g^{ij} (L\cdot M)_{[iL_2,|j|
 M_2]}}{n-(\ell+m-2)}
 \la{3.10.4}
 \ee

Here $L$ stands for $\ell$ antisymmetrised indices, $L_2$ indicates
that the first of these should be omitted and so on. Square brackets
around the multi-indices indicate antisymmetrisation over all of the
indices. The $\d_K$ transformation is generated by the conserved
current $K:=TQ$, where $Q:=Q_{L_2 M_2} D_{+(\ell+m-2)}X^{L_2 M_2}$.
Note that $P$ and $Q$ can be zero and that $N$ is not the Nijenhuis
concomitant except in the special case that $L=M=J$, an almost
complex structure.

The left and right symmetries commute up to the equations of motion.
In the case of $(2,2)$ models, closure of the left and right
algebras separately requires the two type $(1,1)$ tensors $J^\plmi$
to be complex structures. They need not commute unless one demands
off-shell closure without the introduction of further auxiliary
fields. However, any two left and right symmetries of the above type
commute up to a generalised commutator term as a simple argument
shows. Let $\d_{\pm}$ denote left and right variations with two
$L$-tensors, of different rank in general. We have

\begin{equation}
\d_+ \d_- X^i =\d_+\left(a^{-m}L^{\mi i}{}_K D_{-m} X^K\right)\ ,
\label{3.11}
\end{equation}

where $K$ denotes a multi-index with $m$ antisymmetrised indices.
Since all of the $K$ indices are contracted we can replace the
$\d_+$ variation by a covariant variation with the Levi-Civita
connection provided that we take care of the remaining $i$ index.
The explicit connection term drops out in the commutator by
symmetry. In the remaining terms one can introduce either
$\nab^\mi$, acting on $L^\mi$, or $\nab^\pl$, acting on $\d_+ X^k$,
and then show that all of the torsion terms cancel, bar one, again
coming from the $i$ index. However, this cancels in the commutator
too, because the plus and minus connections are swapped in the other
term. One thus finds

\begin{eqnarray}
[\d_+,\d_-]X^i &=&(-1)^n mn\, a^{-m} a^{+n} \left( L^{\mi i}{}_{m
K_2} L^{\pl m}{}_{p L_2} -  L^{\pl i}{}_{m L_2} L^{\mi m}{}_{p
L_2}\right) \times \nn\w2 &\phantom{=}&\left( D_{+(l-1)} X^{L_2}
D_{-(m-1)}X^{K_2}\right)  \times \left(\nab^\pl_-  D_+ X^{p}\right)
\ , \label{3.12}
\end{eqnarray}

the third factor being the equation of motion.  The multi-index $L$
associated with $L^\pl$ stands for $n$ antisymmetrised indices.

\section{Consistency}


In this section we shall examine the consistency of the boundary
conditions, i.e we investigate the orbits of the boundary conditions
under symmetry variations to see if further constraints arise. We
shall show that the supersymmetry variation of the $L$-boundary
condition \eq{3.7} and the $L$-variation of the fermion boundary
condition \eq{2.7} are automatically satisfied if  \eq{3.10} is. To
see this we differentiate \eq{3.10} along $B$ to obtain

\begin{equation}
Y^{(+)}{}_{k,[i_1}{}^{m} \l^{(+)}{}_{i_2\ldots i_{{\ell}+1}]m}=0\ ,
\label{4.1}
\end{equation}

where

\begin{equation}
Y^{(+)}{}_{i,jk}:= (R^{-1})_{jl}( \tnab^{(+)}_i R^l{}_k - H^l{}_{im}
R^m{}_k )\ . \label{4.2}
\end{equation}

Note that we have contracted the derivative with $P$ rather than
$\p$; this is permissible due to the fact that $P\p=\p P=P$.
Equation \eq{4.1} says that $Y^{(+)}$, regarded as a matrix-valued
one-form, takes its values in the Lie algebra of the group which
leaves the form $\l^{(+)}$ invariant. The constraint corresponding
to the superpartner of the $L$-current boundary condition is just
the totally antisymmetric part of \eq{4.1}.

We now consider the variation of the fermionic boundary condition
under $L$-transformations. We need to make both left and right
transformations which together can be written

\begin{eqnarray}
\d X^i &=& 2 a^{+{\ell}} P^i{}_k L^\pl{}^k{}_{j_1\ldots j_{\ell}}
D_{+{\ell}} X^{j_1\ldots j_{\ell}}\w1 &=& 2 a^{-{\ell}} P^i{}_k
L^\mi{}^k{}_{j_1\ldots j_{\ell}} D_{-{\ell}}X^{j_1\ldots j_{\ell}}\
. \label{4.2.1}
\end{eqnarray}

A straightforward computation yields

\begin{equation}
(2\tnab_{[k} R^i{}_{m]}-P^i{}_n H^n{}_{[k|p|} R^p{}_{m]} )
L^{(+)}{}^k{}_{j_1\ldots j_{\ell}}D_{+({\ell}+1)}X^{j_1\ldots
j_{\ell} m} =0 \ . \label{4.3}
\end{equation}

We define

\begin{equation}
Z^{(+)}_{i,jk}= (R^{-1})_{il}(2\tnab_{[j} R^l{}_{k]}+P^l{}_m
H^m{}_{n[j}R^n{}_{k]} )\ , \label{3.24}
\end{equation}

which is the term in the bracket in \eq{4.3} multiplied by $R^{-1}$.
We claim that

\begin{equation}
Y^{(+)}_{i,jk}=Z^{(+)}_{i,jk}\ . \label{3.25}
\end{equation}

This can be proved using  \eq{2.19} with the aid of a little
algebra. Thus we have shown that, if the boundary conditions
\eq{3.10} are consistent, then the constraints following from
supersymmetry variations of the $L$-constraints and from
$L$-variations of the fermionic boundary condition are guaranteed to
be satisfied.

If $\l^\pl=\l^\mi:=\l$ the boundary condition \eq{3.10} typically
implies that $\pm R$ is an element of the group which preserves
$\l$. If this is the case, then \eq{4.1} becomes an identity.
However, it can happen that $R$ is not an element of the invariance
group but that $R^{-1}dR$ still takes its values in the
corresponding Lie algebra. For example, if $\l$ is the two-form of a
$2m$-dimensional K\"ahler manifold and the sign $\h_L\h=-1$, $R$ is
not an element of the unitary group but, since it must have mixed
indices, it is easy to see that $R^{-1}d R$ is itself
$\gu(m)$-valued.

A similar argument applies in the general case, when
$\l^\pl\neq\l^\mi$. In the next section we discuss how the plus and
minus forms are related by an element $V$ of the orthogonal group
(see \eq{5.2.1}). Thus equation \eq{3.10} can be written

\begin{equation}
\l^{(-)}{}_{i_1\ldots i_{{\ell}+1}}=\h_L \h^{\ell}
\l^{(-)}{}_{j_1\ldots j_{{\ell}+1}}\hR^{j_1}{}_{i_1}\ldots
\hR^{j_{{\ell}+1}}{}_{i_{{\ell}+1}}\ , \label{3.10.1}
\end{equation}

where $\hR:=R V^{-1}$. If we differentiate $\eq{3.10.1}$ along the
brane with respect to the minus connection we can then use the above
argument applied to $\hR$.


\section{Target space geometry}


In this section we discuss the geometry of the sigma model target
space in the presence of torsion when the holonomy groups of the
torsion-full connections $\nab^\plmi$ are of special type,
specifically $G_2,\, Spin(7)$ and $SU(3)$. We use only the data
given by the sigma model and use a simple approach based on the fact
that there is a transformation which takes one from one structure to
the other. We begin with $G_2$ and then derive the other two cases
from this by dimensional reduction and oxidation.


\subsection*{$G_2$}


In this case we have a seven-dimensional Riemannian manifold $(M,g)$
with two $G_2$-forms $\vf^\plmi$ which are covariantly constant with
respect to left and right metric connections $\nab^\plmi$ such that
the torsion tensor is $\pm H$. $G_2$ manifolds with torsion have
been studied in the mathematical literature
\cite{Friedrich:2001nh,Friedrich:2001yp} and have arisen in
supergravity solutions \cite{Gauntlett:2002sc}. Bi-$G_2$-structures
have also appeared in this context and have been given an
interpretation in terms of generalised  $G_2$-structures
\cite{Witt:2004vr}. They can be studied in terms of a pair of
covariantly constant spinors from which one can construct the
$G_2$-forms, as well as other forms, as bilinears. We will not make
use of this approach here, preferring to use the tensors given to us
naturally by the sigma model. As noted in \cite{Gauntlett:2002sc}
 there is a common $SU(3)$ structure associated with
the additional forms. We shall derive  this from a slightly
different perspective here.

In most of the literature use is made of the dilatino Killing spinor
equation which restricts the form of $H$. The classical sigma model
does not appear to require this restriction as the dilaton does not
appear until the one-loop level. The dilatino equation is needed in
order to check that one has supersymmetric supergravity solutions
but is not essential for our current purposes.

For $G_2$ there are two covariantly constant forms, the three-form
$\vf$ and its dual four-form $*\vf$ (we shall drop the star when
using indices). The metric can be written in terms of them. A
convenient choice for $\vf$ is

\begin{equation}
\vf=\frac{1}{3!}\vf_{ijk}e^{ijk}=e^{123}-e^1(e^{47}+e^{56})
+e^2(e^{46}-e^{57})-e^3(e^{45}+e^{67}) \label{2.35}
\end{equation}

This form is valid in flat space or in an orthonormal basis, the
$e^i$s being basis forms. Another useful way of think about the
$G_2$ three-form is to write it in a $6+1$ split. We then have

\begin{eqnarray}
\vf_{ijk}&=&\l_{ijk}\nn\w1 \vf_{ij7}&=& \o_{ij}\nn\w1
\vf_{ijk7}&=&-\ghl_{ijk}\ , \label{2.35.02}
\end{eqnarray}

where $i,j,k =1\ldots 6$, and $\{{\l,\ghl,\o}\}$ are the forms
defining an $SU(3)$ structure in six dimensions. The three-forms
$\l$ and $\ghl$ are the real and imaginary parts respectively of a
complex three-form $\O$ which is of type $(3,0)$ with respect to the
almost complex structure defined by $\o$.

On a $G_2$ manifold with skew-symmetric torsion, the latter is
uniquely determined in terms of the Levi-Civita covariant derivative
of $\vf$ \cite{Friedrich:2001nh,Friedrich:2001yp}. This follows from
the covariant constancy of $\vf$ with respect to the torsion-full
connection.

Now suppose we have a bi-$G_2$-structure. The two $G_2$ three-forms
are related to one another by an $SO(7)$ transformation, $V$. If we
start from $\vf^\mi$ this will be determined up to an element of
$G_2^\mi$. So we can choose a representative to be generated by an
element $w\in \gs\go(7)$ of the coset algebra with respect to
$\gg_2^\mi$. This can be written

\begin{equation}
w_{ij}= \vf^\mi_{ijk} v^k \label{5.1}
\end{equation}

and $V=e^w$. The vector $v$ will be specified by a unit vector $N$
and an angle $\a$. It is straightforward to find $V$,

\begin{equation}
V^i{}_j=\cos\a \d^i{}_j + (1-\cos\a) N^i N_j +
\sin\a\,\vf^\mi{}^i{}_{jk} N^k\ . \label{5.2}
\end{equation}

Using

\begin{equation}
\vf^\pl=\vf^\mi V^3\ , \label{5.2.1}
\end{equation}

where one factor of $V$ acts on each of the three indices of $\vf$,
we can find the relation between the two $G_2$ forms explicitly,

\begin{equation}
\vf^\pl_{ijk}=A\vf^\mi_{ijk} + B\vf^\mi_{ijkl} N^l + 3
C\vf^\mi_{[ij}{}^l N_{k]} N_l\ , \label{5.3}
\end{equation}

where

\begin{equation}
A=\cos 3\a,\qquad B=\sin 3\a,\qquad C=1-\cos3\a \ . \label{5.4}
\end{equation}

The dual four-forms are related by

\begin{equation}
\vf^\pl_{ijkl}=(A+C)\vf^\mi_{ijkl} -4 B\vf^\mi_{[ijk} N_{l]} -4
C\vf^\mi_{[ijk}{}^m N_{l]} N_m\ . \label{5.4.1}
\end{equation}

The angle $\a$ is related to the angle between the two covariantly
constant spinors. To simplify life a little we shall follow
\cite{Gauntlett:2002sc} and choose these spinors to  be orthogonal
which amounts to setting $\cos\frac{\a}{2}=0$. We then find

\begin{equation}
\vf^\pl_{ijk}=-\vf^\mi_{ijk} +  6\vf^\mi_{[ij}{}^l N_{k]} N_l\ .
\label{5.5}
\end{equation}

and

\begin{equation}
\vf^\pl_{ijkl}=\vf^\mi_{ijkl} -  8 \vf^\mi_{[ijk}{}^m N_{l]}N_m\ .
\label{5.10}
\end{equation}

We can use the vector $N$ to define an $SU(3)$ structure as above.
We set

\begin{equation}
\o=i_N \vf^\mi\,;\qquad \l=\label{}\vf^\mi-\o\wedge N\,;\qquad
\ghl=i_N*\vf^\mi\ . \label{5.10.1}
\end{equation}

The three-form $\ghl$ is the six-dimensional dual of $\l$ and the
set of forms $\{\o,\l,\ghl\}$ is the usual set of forms associated
with an $SU(3)$ structure in six dimensions. For the plus forms we
have

\begin{eqnarray}
i_N\vf^\pl&=& \o \nn\w1 \vf^\pl-\o\wedge N&=&-\l \nn\w1
i_N*\vf^\pl&=&-\ghl\ . \label{5.10.2}
\end{eqnarray}

Thus a bi-$G_2$-structure is equivalent to a single $G_2$ structure
together with a unit vector (and an angle to be more general). The
unit vector $N$ then allows one to define a set of $SU(3)$ forms as
above. In \cite{Gauntlett:2002sc} it is shown that the projector
onto the six-dimensional subspace is integrable, but this
presupposes that the dilatino Killing spinor equation holds. Since
we make no use of this equation in this paper it need not be the
case that integrability holds.

It is straightforward to construct a covariant derivative $\hnab$
which preserves both $G_2$ structures. This connection has torsion
but this is no longer totally antisymmetric. It is enough to show
that the covariant derivatives of $N$ and $\vf^\mi$ are both zero.
If we write

\begin{equation}
\hnab_i N_j=\nab^\mi_i N_j-S_{i,j}{}^k N_k\ , \label{5.10.3}
\end{equation}

where $S_{i,jk}=-S_{i,kj}$, then these conditions are fulfilled if

\begin{equation}
S_{i,jk}=\half
H_{ijk}+\frac{1}{4}H_i{}^{lm}\vf^\mi_{lmjk}-\frac{3}{2}
H_{ilm}\P^l{}_j\P^m{}_k-\frac{3}{4}H_i{}^{lm}\vf^mi_{jkln} N^n N_m\
. \label{5.10.4}
\end{equation}

Here $\P^i{}_j:=\d^i{}_j-N^i N_j$ is the projector transverse to
$N$.


\subsection*{$SU(3)$}


Manifolds with $SU(3)\times SU(3)$ have arisen in recent studies of
supergravity solutions with flux
\cite{Grana:2004bg}-\cite{Grana:2005ny}. They have also been
discussed in a recent paper on generalised calibrations
\cite{Koerber:2005qi}. A bi-$SU(3)$ structure on a six-dimensional
manifold is given by a pair of a pair of forms
$\{\o^\plmi,\O^\plmi\}$ of the above type which are compatible with
the metric. If the sigma model algebra closes off-shell the complex
structures will be integrable. The transformation relating the two
structures can be found using a similar construction to that used in
the $G_2$ case. However, we can instead derive the relations between
the plus and minus forms by dimensional reduction from $G_2$. To
this end we introduce a unit vector $N'$, which we can take to be in
the seventh direction, and define the $SU(3)$ forms as in equation
\eq{2.35.02} above. We consider only the simplified
bi-$G_2$-structure and we also then take the unit vector $N$ to lie
within the six-dimensional space. The unit vector $N$ now defines an
$SO(6)$ transformation. The relations between the plus and minus
forms are given by

\begin{eqnarray}
\o^\pl_{ij}&=& -\o^\mi_{ij} + 4 \o^\mi_{[i}{}^k N_{j]} N_k \nn \w1
\l^\pl_{ijk}&=& -\l^\mi_{ijk} +6\l_{[ij}{}^l N_{k]} N_l. \nn \w1
\ghl^\pl_{ijk}&=& \ghl^\mi_{ijk}- 6\ghl^\mi_{[ij}{}^l N_{k]} N_l\ .
\label{5.4.3}
\end{eqnarray}

We can rewrite this in complex notation if we introduce the
three-forms $\O^\plmi:=\l^{\plmi}+i\ghl^{\plmi}$ and split $N$ into
$(1,0)$ and $(0,1)$ parts, $n,\bar n$. So

\begin{equation}
N_i=n_i + \bar n_i\,;\qquad i\o_{ij} N^j =n_i-\bar n_i \ .
\label{5.4.4}
\end{equation}

Note that $n\cdot\bar n=\frac{1}{2}$. Then equations \eq{5.4.3} are
equivalent to

\begin{eqnarray}
\o^\pl_{ij}&=& -\o^\mi_{ij} -2in_{[i} \bar n_{j]}\nn\w1
\O^\pl_{ijk}&=& 6\bar\O^\mi_{[ij}{}^l n_{k]} n_l\ . \label{5.4.5}
\end{eqnarray}

This type of bi-$SU(3)$-structure is therefore equivalent to a
single $SU(3)$ structure together with a normalised $(1,0)$-form.


\subsection*{$Spin(7)$}


A $Spin(7)$ structure on an eight-dimensional Riemannian manifold is
specified by a self-dual four-form $\F$ of a certain type. In a
given basis its components can be constructed from those of the
$G_2$ three-form. Thus

\begin{equation}
\F_{abcd}=\vf_{abcd}\,;\qquad {\rm and} \qquad\F_{abc8}=\vf_{abc}\ ,
\label{5.4.6}
\end{equation}

where, in this section, $a,b,\ldots$ run from 1 to 7 and
$i,j,\ldots$ run from 1 to 8.  $Spin(7)$ geometry with
skew-symmetric torsion has been discussed
\cite{Friedrich:2001nh,Ivanov:2001ma} and generalised $Spin(7)$
structures have also been studied \cite{Witt:2004vr}. A
bi-$Spin(7)$-structure on a Riemannian manifold consists of a pair
of such forms, covariantly constant with respect to $\nab^\plmi$. We
can again get from the minus form to the plus form by an orthogonal
transformation, but since the dimension of $SO(8)$ minus the
dimension of $Spin(7)$ is seven it is described by seven parameters.
In the presence of a $Spin(7)$ structure one of the chiral spinor
spaces, $\D_s$, say, splits into one- and seven-dimensional
subspaces, $\D_s=\bbR\oplus \D_7$. The transformation we seek will
be described by a unit vector $n^a\in \D_7$ together with an angle.

It will be useful to introduce some invariant tensors for $Spin(7)$
using this decomposition of the spin space. We set

\begin{eqnarray}
\f_{ajk}&=&\cases{\f_{abc}=\vf_{abc} \cr \f_{ab8}=\d_{ab}\cr}
\label{5.4.7}\w2
\f_{abkl}&=&\cases{\f_{ab}{}^{cd}=\vf_{ab}{}^{cd}-2\d_{[ab]}^{cd}
\cr \f_{abc8}=-\vf_{abc}\cr}\ , \label{5.4.8}
\end{eqnarray}

where $\vf_{abc}$ is the $G_2$ invariant. It will also be useful to
define

\begin{equation}
\f_{aijkl}:=\f_{ab[ij}\f^b{}_{kl]}\ . \label{5.4.9}
\end{equation}

The $Spin(7)$ form itself can be written as

\begin{equation}
\F_{ijkl}=\f_{a[ij}\f^a{}_{kl]}\ . \label{5.4.10}
\end{equation}

The space of two-forms splits into $7+21$, and one can project onto
the seven-dimensional subspace by means of $\f_{ajk}$.  With these
definitions we can now oxidise the $G_2$ equations relating the plus
and minus structure forms to obtain

\begin{equation}
\F^\pl_{ijkl}= -\F^\mi_{ijkl}-6 n_a n_b \f^{\mi a}{}_{[ij}\f^{\mi
b}{}_{kl]} \label{5.4.11}\ .
\end{equation}

Here the unit vector $N$ in the $G_2$ case becomes the unit spinor
$n$.


\section{Examples of solutions}


In this section we look at solutions to the boundary conditions for
the additional symmetries which can be identified with various types
of brane.  We shall go briefly through the main examples, confining
ourselves to $U(\frac{n}{2})$, $SU(\frac{n}{2})$ and the exceptional
cases $G_2$ and $Spin(7)$.


\subsection*{$U(\frac{n}{2})\equiv U(m)$}


This case corresponds to $N=2$ supersymmetry. For $H=F=0$ we assume
that the supersymmetry algebra closes off-shell so that $M$ is a
K\"ahler manifold with complex structure $J$, hermitian metric $g$
and K\"ahler form $\o$. The K\"ahler form is closed and covariantly
constant. The boundary conditions for the second supersymmetry,
which can be viewed as an additional symmetry with $\l=\o$, imply

\begin{equation}
\o_{ij}=\pm \o_{kl} R^k{}_i R^l{}_j\ . \label{2.28}
\end{equation}

Thus there are two possibilities, type A where $JR=-RJ$ and type B
where $JR=RJ$ \cite{Ooguri:1996ck}.  Consider type B first. In this
case the brane inherits a K\"ahler structure from the target space
and so has dimension $2k$. If there is a non-vanishing gauge field
$F$, it must be of type $(1,1)$ with respect to this structure. The
calibration form is $\o^k$.

For type B with zero $F$ field, $J$ is off-diagonal in the
orthonormal basis in which $R$ takes its canonical form

\begin{equation}
R=\left(\begin{array}{cc} 1_p & 0\w2 0 & -1_q
\end{array}\right)\ ,
\label{2.29}
\end{equation}

where $p$ and $q$ denote the dimensions of $B$ and the transverse
tangent space, $p+q=n$, and $1_p,\,1_q$ denote the corresponding
unit matrices. The only possibility is $p=q=m$. The K\"ahler form
vanishes on both the tangent and normal bundles to the brane, so
that the brane is Lagrangian.

When the $F$ field is non-zero the situation is more complicated. We
may take $R$ to have the same block-diagonal form as in \eq{2.29}
but with $1_p$ replaced by $R_p$. From \eq{2.12.4}

\begin{equation}
R_p= (1+F)^{-1} (1-F) \label{2.29.1}
\end{equation}

The analysis of $JR=-RJ$ shows that the brane is coisotropic
\cite{Kapustin:2001ij}. This means that there is a $4k$-dimensional
subspace in each tangent space to the brane where $J$ is
non-singular, there is an $r$-dimensional subspace on which it
vanishes, and the dimension of the normal bundle is also $r$.  The
product  $(J_p F)$ is an almost complex structure and both $J_p$ and
$F$ are of type $(2,0)+(0,2)$ with respect to $(J_p F)$. For $m=3$
we can therefore only have $p=5$. For $m=4$ we can have $p=5$ but we
can also have a space-filling brane with $p=8$.

$N=2$ sigma models with boundary and torsion have been discussed in
\cite{Lindstrom:2002jb}; the geometry associated with the boundary
conditions is related to generalised complex geometry
\cite{Zabzine:2004dp,Kapustin:2004gv}.


\subsection*{$SU(\frac{n}{2})\equiv SU(m)$}


In the Calabi-Yau case we have, in addition to the K\"ahler
structure, a covariantly constant holomorphic $(m,0)$ form $\O$
where $m= \frac{n}{2}$. There are two independent real covariantly
constant forms, $\l$ and $\ghl$, which can be taken to be the real
and imaginary parts of $\O$. The corresponding $L$-tensors which
define the symmetry transformations are related by

\begin{equation}
\hL^i{}_{j_1\ldots j_{m-1}}= J^i{}_k L^k{}_{j_1\ldots j_{m-1}}
\label{2.33}
\end{equation}

Because there are now two currents we can introduce a phase rather
than a sign in the boundary condition. Thus

\begin{equation}
\O_{i_1\ldots i_m} = e^{i\a} \O_{j_1\ldots j_m}
R^{j_1}{}_{i_1}\ldots R^{j_m}{}_{i_m} \label{2.33.1}
\end{equation}

A second possibility is that $\O$ on the right-hand side is replaced
by $\bar \O$. For type B branes, the displayed equation is the
correct condition. The $R$-matrix is the sum of holomorphic and
anti-holomorphic parts, $R=\cR\oplus\bar{\cR}$, and \eq{2.33.1}
implies that

\begin{equation}
\det \cR=e^{i\a}\ . \label{2.33.2}
\end{equation}

If $F=0$ this fixes the phase, but if $F\neq 0$ it imposes a
constraint on $F$ which must in any case be a $(1,1)$ form (from
$JR=RJ$) \cite{Kapustin:2003se}. The constraint is

\begin{equation}
\det\cR_p=e^{i\a} (-1)^{\frac{q}{2}}\ , \label{2.33.3}
\end{equation}

or

\begin{equation}
\det (1+f)= e^{i\a} (-1)^{\frac{q}{2}} \det(1-f) \label{2.33.4}
\end{equation}

where $f^a{}_b=g^{a\bar c} F_{\bar c b}$, in a unitary basis.

For type A branes, from $JR=-RJ$ it follows that $R$ maps
holomorphic vectors to anti-holomorphic ones and vice versa so that
$\bar \O$ must be used in \eq{2.33.1}. In the case that $F=0$ the
brane is a SLAG with $\gR\ge\O$ as the calibration form. For $F\neq
0$ we have  coisotropic branes with an additional constraint on the
gauge field \cite{Kapustin:2003se}.

The geometry of the bi-$SU(m)$ case has been studied from the point
of view of generalised geometry and generalised calibrations in
\cite{Koerber:2005qi}.


\subsection*{$G_2$}


The boundary conditions associated with the $G_2$ currents are

\begin{eqnarray}
\vf &=& \h_L \vf R^3 \nn \w1 *\vf &=& \h_L *\vf R^4\, \det R
\label{2.34}
\end{eqnarray}

We consider first $F=H=0$. From the first of these equations it
follows that $(\h_L R)\in G_2$. From this it follows that the sign
in the boundary condition for $*\vf$ is always positive because the
sign of the determinant of $R$ is equal to $\h_L$. Thus the second
constraint reduces to $*\vf=*\vf R^4$.

There are two possibilities depending on the sign of $\h_L$. If it
is positive then non-zero components of $\vf$ must have an even
number of normal indices, whereas if it is negative they must have
an odd number of non-zero components. Since $(\h_L R)\in G_2$, and
is symmetric, it can be diagonalised by a $G_2$ matrix so that we
can bring $R$ to its canonical form in a $G_2$ basis. Looking at the
components of $\vf$ we see that the only possibilities which are
compatible with the preservation of the non-linear symmetries on the
boundary are either $\h_L=1$, in which case $B$ is a
three-dimensional associative cycle, or $\h_L=-1$ in which case $B$
is a four-dimensional co-associative cycle \cite{Becker:1996ay}.

Now let us turn to $F\neq 0$, but $H=0$. We shall assume that the
tangent bundle $M$, restricted to the brane, splits into three,
$TM|_B=T_1\oplus T_2\oplus N$, of dimensions $p_1,p_2$ and $q$
respectively. $N$ is the normal bundle and $R|_{T_2}=1_{p_2}$. If
there is at least one normal direction we may assume that one of
these is $7$ in the conventions of \eq{2.35.02}. Thus the problem is
reduced to a six-dimensional one, at least algebraically. The
six-dimensional boundary conditions are (where $R$ is now a $6\xz 6$
matrix),

\begin{eqnarray}
\l&=&\h_L\l R^3\nn\w1 \ghl&=&\h_L \ghl R^3 \det R \nn\w1
\o&=&-\h_L\o R^2 \ , \label{2.35.1}
\end{eqnarray}

in an obvious notation. If the sign is negative the brane is type B,
whereas if $\h_L=+1$  we have type A. These are the same conditions
as we have just discussed in the preceding section, the only
difference being that the phase is not arbitrary. The constraints on
the $F$ field are therefore slightly stronger.

The last possibility is a space-filling brane in seven dimensions.
Since $F$ is antisymmetric there must be at least one trivial
direction for $R$ so that we can again reduce the algebra to the
six-dimensional case. The only possibilty is $\h=+1$ in which case
we have type B. The non-trivial dimension must be even, and since
$\det{\cR}=1$ the case $p=2$ is also trivial.

Now let us consider the case with torsion. The boundary condition
for the non-linear symmetries associated with the forms yield

\begin{eqnarray}
\vf^{\pl}&=& \h_L\vf^{\mi} R^3\nn\w1 *\vf^{\pl}&=&\h_L *\vf^{\mi}
R^4\det R\ , \label{5.13}
\end{eqnarray}

When the brane is normal to $N$ we find, on the six-dimensional
subspace,

\begin{eqnarray}
\l&=&-\h\l R^3\nn\w1 \ghl&=&-\h \ghl R^3 \det R \nn\w1 \o&=&-\h\o
R^2 \ , \label{5.14}
\end{eqnarray}

The analysis is very similar to the case of zero torsion with $F$.
One finds that $\h_L=-1$ corresponds to type B while $\h_L=+1$ is
type A.  In particular, for type B there is a five-brane which
corresponds to the five-brane wrapped on a three-cycle discussed in
the supergravity literature
\cite{Gauntlett:2001ur,Gauntlett:2002sc}.


\subsection*{$Spin(7)$}


In the absence of torsion, the boundary condition associated with
the conserved current is

\begin{equation}
\F=\ghh \F R^4\ , \label{2.42}
\end{equation}

for some sign factor $\ghh$. If this is negative then $\det R$ is
also negative so that the dimension of $B$ must be odd. Furthermore,
$\F$ must have an odd number of normal indices with respect to the
decomposition of the tangent space induced by the brane. However,
one can show that such a decomposition is not compatible with the
algebraic properties of $\F$. Therefore the sign $\ghh$ must be
positive. It is easy to see that a four-dimensional $B$ is
compatible with this, and indeed we then have the standard Cayley
calibration with $\F$ pulled-back to the brane being equal to the
induced volume form. On the other hand if $B$ has either two or six
dimensions one can show that it is not compatible with the $Spin(7)$
structure. As one would expect, therefore, the only brane compatible
with the non-linear symmetry associated with $\F$ on the boundary is
the Cayley cycle \cite{Becker:1996ay}.

If $F\neq0$, but $H=0$, and if we assume that there is at least one
direction normal to the brane, then the $Spin(7)$ case reduces to
$G_2$ (with $F\neq0$). If the brane is space-filling but there is at
least one trivial direction, then there must be at least two by
symmetry and again we recover the $G_2$ case. But we can also have a
space-filling brane which is non-trivial in all eight directions.

\section*{Acknowledgements}

This work was supported in part by EU grant (Superstring theory)
MRTN-2004-512194, PPARC grant number PPA/G/O/2002/00475 and VR grant
621-2003-3454. PSH thanks the Wenner-Gren foundation.


\end{document}